# The Intense Starburst HDF850.1 in a Galaxy Overdensity at z=5.2 in the Hubble Deep Field

*Fabian Walter[1,2], Roberto Decarli[1], Chris Carilli[2,3], Frank Bertoldi[4], Pierre Cox[5], Elisabete Da Cunha[1], Emanuele Daddi[6], Mark Dickinson[7], Dennis Downes[5], David Elbaz[6], Richard Ellis[8], Jacqueline Hodge[1], Roberto Neri[5], Dominik Riechers[8], Axel Weiss[9], Eric Bell[10], Helmut Dannerbauer[11], Melanie Krips[5], Mark Krumholz[12], Lindley Lentati[3], Roberto Maiolino[13], Karl Menten[9], Hans-Walter Rix[1], Brant Robertson[14], Hyron Spinrad[15], Dan Stark[14], Daniel Stern[16]*

[1] Max-Planck Institut für Astronomie, Königstuhl 17, D-69117, Heidelberg, Germany.

[2] National Radio Astronomy Observatory, Pete V. Domenici Array Science Center, P.O. Box O, Socorro, NM, 87801, USA.

[3] Cavendish Laboratory, J J Thomson Avenue, Cambridge University, Cambridge CB3 0HE, UK

[4] Argelander Institute for Astronomy, University of Bonn, Auf dem Hügel 71, 53121 Bonn, Germany.

[5] IRAM, 300 rue de la piscine, F-38406 Saint-Martin d'Hères, France.

[6] Laboratoire AIM, CEA/DSM-CNRS-Université Paris Diderot, Irfu/Service d'Astrophysique, CEA Saclay, Orme des Merisiers, 91191 Gif-sur-Yvette Cedex, France.

[7] National Optical Astronomy Observatory, 950 North Cherry Avenue, Tucson, AZ 85719, USA.

[8] Astronomy Department, California Institute of Technology, MC105-24, Pasadena, CA 91125, USA.

[9] Max-Planck-Institut für Radioastronomie, Auf dem Hügel 69, 53121 Bonn, Germany.

[10] Department of Astronomy, University of Michigan, 500 Church Street, Ann Arbor, MI 48109, USA.

[11] Universität Wien, Institut für Astronomie, Türkenschanzstraße 17, 1080 Wien, Austria

[12] Department of Astronomy and Astrophysics, University of California, Santa Cruz, CA 95064, USA.

[13] INAF-Osservatorio Astronomico di Roma, via di Frascati 33, 00040 Monte Porzio Catone, Italy.

[14] Department of Astronomy, University of Arizona, 933 North Cherry Avenue, Tucson, AZ 85721, USA.

[15] Department of Astronomy, University of California at Berkeley, Berkeley, CA 94720, USA.

[16] Jet Propulsion Laboratory, California Institute of Technology, 4800 Oak Grove Drive, Pasadena, CA 91109, USA.

**The Hubble Deep Field (HDF) is a region in the sky that provides one of the deepest multi-wavelength views of the distant universe and has led to the detection of thousands of galaxies seen throughout cosmic time[1]. An early map of the HDF at a wavelength of 850 microns that is sensitive to dust emission powered by star formation revealed the brightest source in the field, dubbed HDF850.1[2]. For more than a decade, this source remained elusive and, despite significant efforts, no counterpart at shorter wavelengths, and thus no redshift, size or mass, could be identified[3-7]. Here we report, using a millimeter wave molecular line scan, an unambiguous redshift determination for HDF850.1 of z=5.183. This places HDF850.1 in a galaxy overdensity at z~5.2 in the HDF, corresponding to a cosmic age of only 1.1 Gyr after the Big Bang. This redshift is significantly higher than earlier estimates[3,4,6,8] and higher than most of the >100 sub-millimeter bright galaxies identified to date. The source has a star formation rate of 850 $M_{sun}$ yr$^{-1}$ and is spatially resolved on scales of 5 kpc, with an implied dynamical mass of ~$1.3\times10^{11}$ $M_{sun}$, a significant fraction of which is present in the form of molecular gas. Despite our accurate redshift and position, a counterpart arising from starlight remains elusive.**

We have obtained a full frequency scan of the 3 mm band towards the HDF using the IRAM Plateau de Bure Interferometer (PdBI). The observations covered the frequency range from 80-115 GHz in 10 frequency settings at uniform sensitivity and at a resolution (~2.3") that is a good match to galaxy sizes at high redshift. They resulted in the detection of two lines of Carbon Monoxide (CO), the most common tracer for molecular gas at high redshift[9], at 93.20 GHz and 111.84 GHz at the position of HDF850.1. Identifying these lines with the J=5 and J=6 rotational transitions of CO gives a redshift for HDF850.1 of z=5.183. This redshift was then unambiguously confirmed by the PdBI detection of the 158 μm line of ionized carbon ([CII], redshifted to 307.38 GHz), one of the main cooling lines of the star-forming interstellar medium. Stacking of other molecules covered by our frequency scan that trace higher volume densities did not lead to a detection (see Supplementary Information). Subsequently, the J=2 line of CO has also been detected using the NRAO Jansky Very Large Array (Jansky VLA) at 37.29 GHz. The observed [CII] and CO spectra towards HDF850.1 are shown in Fig. 1.

The beamsize of our CO observations (~2.3", 15 kpc at z=5.183) is too large to spatially resolve the molecular gas emission in HDF850.1. However the [CII] and underlying continuum observations (~1.2" x 0.8") show that the source is extended (hitherto, the interstellar medium has been spatially resolved only in extremely rare quasar host galaxies at such high redshift[10]). A single Gaussian fit yields a deconvolved size of 0.9±0.3", or 5.7±1.9 kpc at the redshift of the source. Fig. 2 shows the maps of total [CII] emission (left) as well as the red- and blue-shifted parts of the [CII] line (right) superposed on the deepest available Hubble Space Telescope (HST) images of the HDF[1]. The derived dynamical mass is $M_{dyn}$~1.3±0.4 x $10^{11}$ $M_{sun}$ assuming an arbitrary inclination of 30 degrees. An alternative interpretation is that the source is a merger of two galaxies, rather than a single rotating disk, which would lower the implied dynamical mass. Fig. 2 shows that the source is completely obscured in the observed optical and near-infrared wavebands (i.e. the rest-frame UV). There is no indication of HDF850.1 harboring an active galactic nucleus powered by a supermassive black hole (quasar)[11].

The CO(6-5)/CO(2-1) line luminosity ratio (in units of K km s$^{-1}$ pc$^2$)[9] is 0.23±0.05. Assuming that the gas is being emitted from the same volume, this implies that the high-J CO emission is sub-thermally excited on galactic scales, less than seen in the nuclei of local starburst galaxies[12]. Using a standard large velocity gradient (LVG) model we find that the observed CO line intensities can be fit with a moderate molecular hydrogen density of $10^{3.2}$ cm$^{-3}$ and a kinetic temperature of 45 K for virialized clouds (dv/dr=1.2 km s$^{-1}$ pc$^{-1}$). We caution though that these numbers would change if the CO transitions were not emitted from the same volume. The predicted CO(1-0) line luminosity is 4.3 x $10^{10}$ K km s$^{-1}$ pc$^2$, close to the measured value for CO(2-1). Depending on the choice of α, the CO-to-H$_2$ conversion factor, this line luminosity implies a molecular gas mass of $M_{H2}$ = 3.5 x (α/0.8) x $10^{10}$ $M_{sun}$; here α=0.8, in units of $M_{sun}$ (K km s$^{-1}$ pc$^2$)$^{-1}$, is the conversion factor adopted for ultra-luminous infrared galaxies (ULIRGs)[13] and thought to be applicable to sub-millimeter bright objects[14]. The implied molecular gas mass fraction is $M_{H2}/M_{dyn}$ ~ 0.25±0.08 (α/0.8); i.e. even with a low ULIRG conversion factor the molecular gas constitutes a significant fraction of the overall dynamical mass. This molecular gas mass (and fraction) is comparable to what is found in other sub-millimeter bright galaxies that are typically located at much lower redshift[14,15].

The line-free channels of the observations (Fig. 1) were used to constrain the underlying continuum emission. Our accurate position of the rest-frame 158 μm emission is indicated as a cross in Fig. 2 (right). We combine our continuum detections at 307 GHz and 112 GHz with published values and new Herschel Space Telescope observations to constrain the far-infrared (FIR) properties of the source (see Supplementary Information for details). Our best fit gives a FIR luminosity of $L_{FIR}$=6.5±1 x $10^{12}$ $L_{sun}$, a dust temperature of 35±5 K (i.e., broadly consistent with the average kinetic temperature of the molecular gas), a dust mass of $M_{dust}$ = 2.75±0.5 x $10^8$ $M_{sun}$ and a star formation rate of 850 $M_{sun}$ yr$^{-1}$ (with an uncertainty of ~30%). Given the extent of the source this results in an galaxy-averaged star formation rate surface density of 850 $M_{sun}$ yr$^{-1}$ / (π x (2.8 kpc)$^2$) ~ 35 $M_{sun}$ yr$^{-1}$ kpc$^{-2}$ (uncertainty ~50%), more than an order of magnitude less than found in nearby merging systems and a compact quasar host galaxy at z=6.42 that has been studied in similar detail[10]. HDF850.1 falls on the universal local star formation law that relates the average surface density of the star formation rate to that of the molecular gas mass per local free-fall time[16]. The estimated surface density would increase if future observations resolved the source structure.

The resulting [CII]/FIR luminosity ratio of $L_{[CII]}/L_{FIR}$ = 1.7 ± 0.5 x $10^{-3}$ in HDF850.1 is comparable to what is found in normal local star-forming galaxies[17], but is an order of magnitude higher than what is found in a z=6.42 quasar[10], the only other high-z system where the [CII] emission could be resolved to date. Recent studies indicate that this ratio is a function of environment, with a low value ($L_{[CII]}/L_{FIR}$ ~ 1 x $10^{-4}$) for luminous systems dominated by a central black hole (quasars) and a high ratio (up to $L_{[CII]}/L_{FIR} \sim 1 \times 10^{-2}$) for low-metallicity environments. Our relatively high ratio in $L_{[CII]}/L_{FIR}$ is consistent with HDF850.1 being a high redshift star-forming system in a non-quasar environment[17].

An inspection of the distribution of galaxies towards HDF850.1 that have spectroscopic redshifts shows that there is an overdensity of galaxies at the exact redshift of HDF850.1, including a quasar at z=5.186[18] (Fig. 3 and Supplementary Information). This makes this region one of the most distant galaxy overdensities known to date[19]. An elliptical galaxy at z=1.224[20] that is situated close to HDF850.1 in projection (~1" to the NE) could potentially act as a gravitational lens for this source[3,4,21]. Using a velocity dispersion of 146 km s$^{-1}$ in a singular isothermal sphere for this elliptical galaxy[4] and our new redshift and position of HDF850.1, we derive an amplification factor of ~1.4. A similar flux amplification is found for a simple point source lens model with mass $3.5 \times 10^{11}$ $M_{sun}$. This implies that even if lensing is occurring, the quantities derived here would not need to be revised significantly.

HDF850.1 remains outstanding in the study of dust-obscured starbursts at high redshift, being one of the first such sources discovered, and yet evading detection in the optical and near-infrared. Its redshift of z=5.183 enforces the presence of a high redshift tail (z>4) of sub-millimeter bright star-forming (non-AGN/quasar) galaxies (currently there are only about half a dozen systems known)[22-26]. Only a small fraction of sub-millimeter bright sources is expected to be at very high redshift[27] --- it is thus ironic that the first blank-field source belongs to this subgroup. HDF850.1's large spatial extent, in combination with the modest CO excitation, a moderate surface density of its star formation rate, and a high [CII]/FIR luminosity ratio all point to the presence of a spatially extended major starburst that is completely obscured even in the deepest HST images available for the HDF. The absence of a possible counterpart in the available deep imaging, even though the star-forming interstellar medium is distributed over many square kpc, makes this source extreme[22-24]. Given its high molecular gas mass ($3.5 \times (\alpha/0.8) \times 10^{10}$ $M_{sun}$) and star formation rate (850 $M_{sun}$ yr$^{-1}$) HDF850.1 can build a significant stellar component as early as z~4[28] (~few hundred Myr from z~5). Blind line searches through spectral scans at millimeter wavelengths as performed here thus play a fundamental role in unveiling the nature of star-forming galaxies that are completely obscured in the (restframe) optical and UV even if multi-wavelength data at unparalleled depth are available.

**Acknowledgements:** This work is based on observations carried out with the IRAM Plateau de Bure Interferometer. IRAM is supported by MPG (Germany), INSU/CNRS (France) and IGN (Spain). The Jansky Very Large Array of the National Radio Astronomy Observatory (NRAO) is a facility of the National Science Foundation operated under cooperative agreement by Associated Universities, Inc. DR acknowledges support from NASA through a Spitzer Space Telescope grant. RD acknowledges funding through DLR project FKZ 50OR1004.

**Contributions:** F.W. had the overall lead of the project. The PdBI data were analyzed by R.D., F.W., P.C., R.N., M.K. and D.D. The Jansky VLA data reduction was performed by C.C., J.H., and L.L. The molecular gas excitation was lead by A.W. Spectroscopic redshift information was provided by M.D., R.E., H.S., D.S. and D.P.S. The SED analysis including new Herschel data was lead by E.dC, D.E. and E.D. An updated lensing model was provided by D.D. All authors helped with the proposal, data analysis and interpretation.

**Competing Interest Statement:** The authors declare that they have no competing financial interests.

**Corresponding authors:** Correspondence and requests for material should be addressed to F.W. (walter@mpia.de).

**Figures**

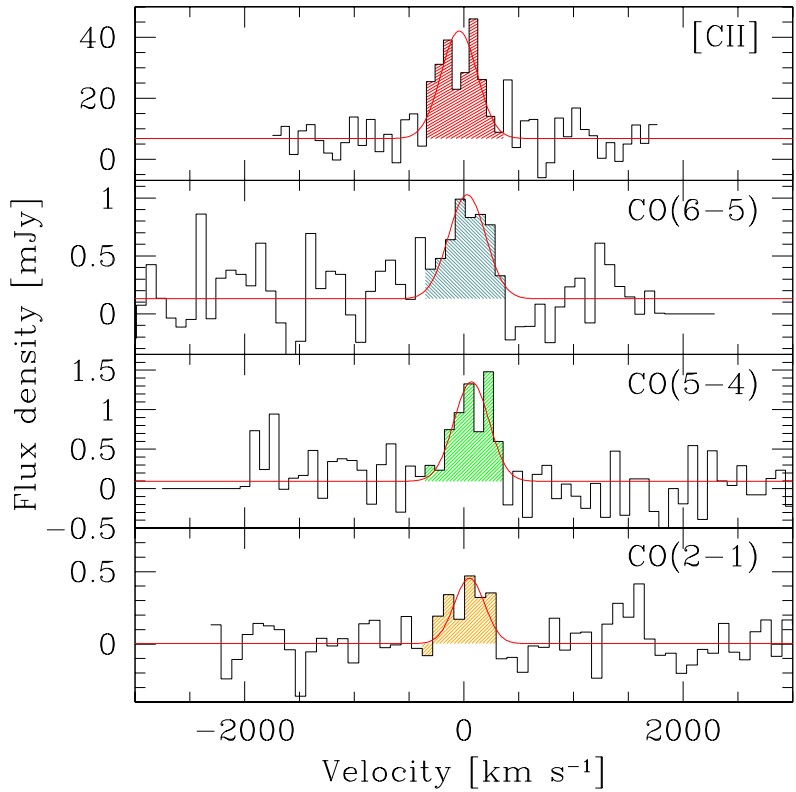

**Figure 1:** Detection of four lines tracing the star-forming interstellar medium in HDF850.1. From top to bottom: [CII], $\nu_{obs}$ = 307.383 GHz; CO(6-5), $\nu_{obs}$ = 111.835 GHz; CO(5-4), $\nu_{obs}$ = 93.202 GHz, CO(2-1), $\nu_{obs}$ = 37.286 GHz. Zero velocity corresponds to a redshift of z=5.183. Continuum emission is detected in the top two panels at 6.80±0.8 mJy and 0.13±0.03 mJy, respectively. We derive a 3σ continuum limit of 30 μJy from the Jansky VLA observations at 37.3 GHz using a bandwidth larger than shown here. Gaussian fits to the lines give a full width at half maximum (FWHM) of 400±30 km s$^{-1}$, narrower than typically found in sub-millimeter selected galaxies[13]. The observed integrated line flux densities are: S[CII]=14.6±0.3 Jy km s$^{-1}$, S[CO(6-5)]=0.39±0.1 Jy km s$^{-1}$, S[CO(5-4)]=0.50±0.1 Jy km s$^{-1}$ and S[CO(2-1)]=0.17±0.04 Jy km s$^{-1}$. The resulting line luminosities are[9]: 5.0, 1.0, 1.9 and 4.1 x 10$^{10}$ K km s$^{-1}$ pc$^2$ or 1,104, 10.6, 11.4 and 1.5 x 10$^7$ L$_{sun}$ (uncertainties as given for integrated line flux densities). LVG modeling gives a predicted CO(1-0) line luminosity of 4.3 x 10$^{10}$ K km s$^{-1}$ pc$^2$.

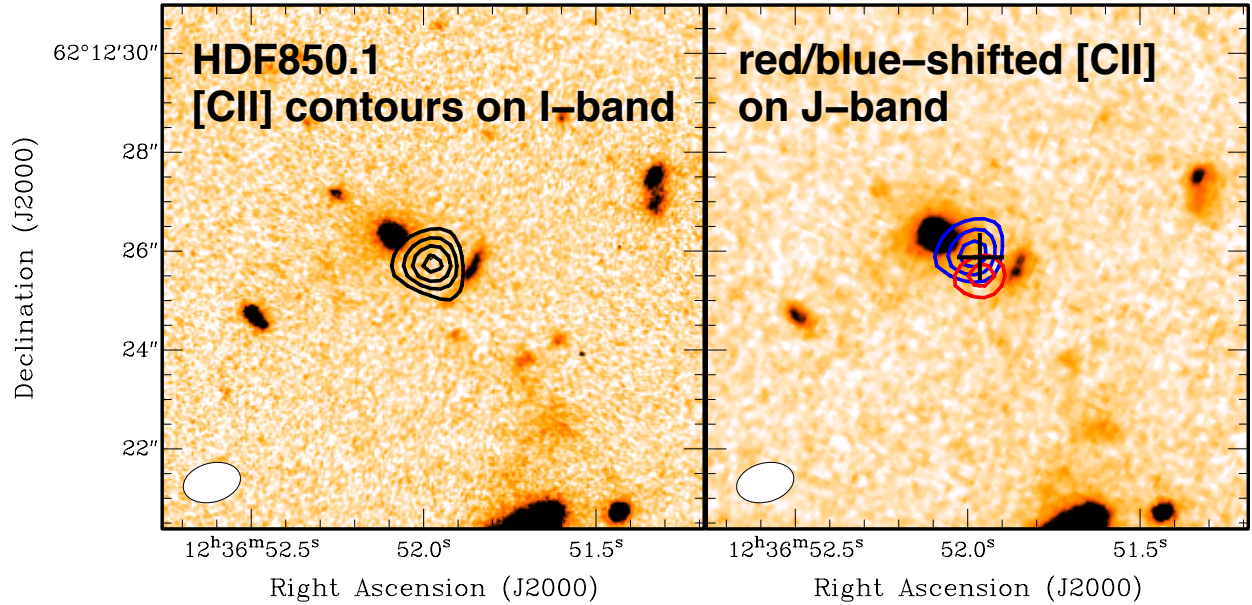

**Figure 2:** [CII] line emission towards HDF850.1. *Left:* [CII] contours on top of a deep HST image[1] of the region in a filter (I band) that covers the Ly-α line and UV continuum at z=5.183. [CII] contours show the averaged emission over 700 km s$^{-1}$ and are plotted at 5, 7, 9, 11 mJy beam$^{-1}$ (1σ=1.3 mJy beam$^{-1}$). A Gaussian fit to the emission gives a deconvolved size of 0.9±0.3" or 5.7±1.9 kpc at z=5.183. The underlying continuum emission (not shown) is also extended on the same scales. *Right:* The blue and red contours indicate the approaching and receding [CII] emission relative to the systemic redshift of z=5.183. The color shows a deep HST image in a longer wavelength filter (NICMOS J band)[29]. The cross indicates the position and its 5σ uncertainty of the rest-frame 158 μm continuum emission peak (RA: 12h36m51.976s, DEC: 62°12'25.80" in the J2000.0 system), consistent with earlier millimeter interferometric measurements[3,6] at lower resolution. The [CII] contours have been derived by averaging the spectrum (Fig. 1) from -400 km s$^{-1}$ to 0 km s$^{-1}$ and 0 km s$^{-1}$ to +400 km s$^{-1}$ and are plotted at levels of 7, 10 and 13 mJy beam$^{-1}$ (1σ=1.8 mJy beam$^{-1}$), respectively. In both panels the beamsize of the [CII] observations (1.23" x 0.81") is indicated in the bottom left corner. From the spatial offset (total offset=0.9", i.e. radius: 0.45", or r=2.8 kpc) and the FWHM of the line, we derive an approximate dynamical mass of $M_{dyn} \sim 3.4 \times 10^{10}$ $M_{sun}$ / (sin $i$)$^2$ where $i$ is the (unknown) inclination of the system (using $M_{dyn}$ sin$^2 i$ = 1.3 x (FWHM/2)$^2$ r/G, where G is the gravitational constant[30]). These deep HST images of the HDF fail to reveal the (rest-frame) UV/optical counterpart of the galaxy that is forming stars at a rate of ~850 $M_{sun}$ yr$^{-1}$.

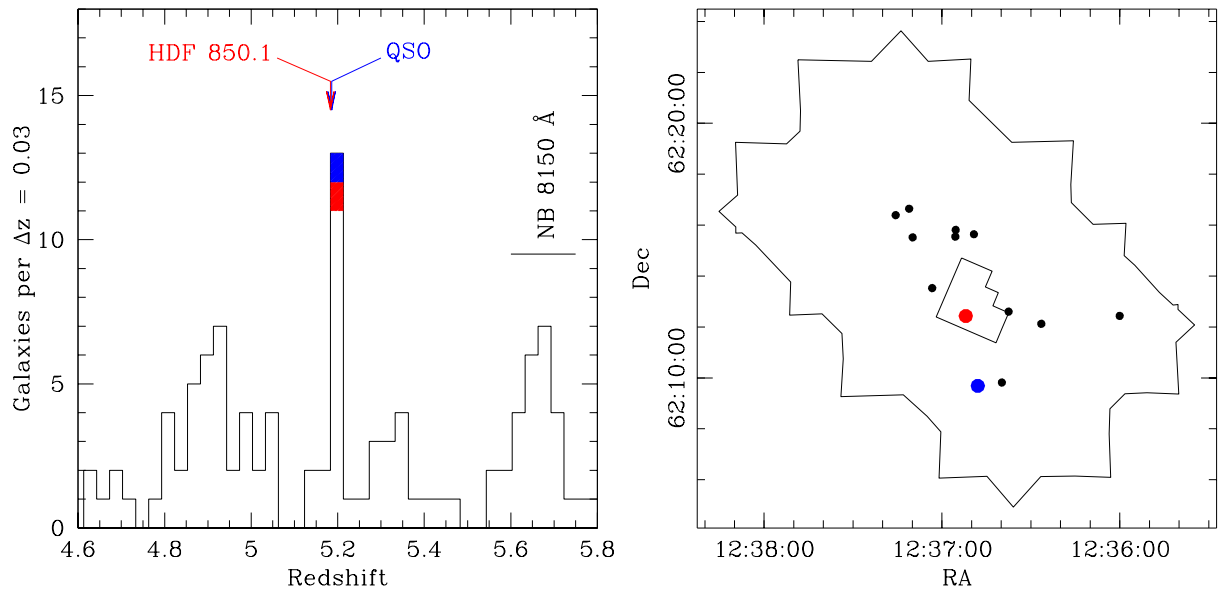

**Figure 3:** Distribution of galaxies near HDF850.1 *Left:* Distribution of spectroscopic redshifts towards the HDF and its surrounding (GOODS-N). HDF850.1 is indicated by a red color, the quasar at the same redshift[18] is shown in blue. There is an overdensity of galaxies in the redshift bin that contains HDF850.1. The high source density at z~5.7 is an observational artefact due to narrow-band Lyman-α imaging surveys of the region (with spectroscopic follow-up) that are sensitive to this particular narrow redshift range. *Right:* Spatial coverage of the sources in the redshift bin z=5.183-5.213. The small border indicates the size of the HDF – the larger border shows the surrounding area of GOODS-N. The presence of a strongly star-forming galaxy (HDF850.1) and a quasar[18] in this region provides evidence for cosmic structure formation in the first Gyr of the universe. See Supplemental Information for more details.

# Supplementary Information

## 1. Details on the Galaxy Overdensity at z=5.2

The redshifts of the galaxies shown in the left panel of Figure 3 have been compiled from a number of published spectroscopic surveys of the GOODS-N region[20,31-38] and include new data from ongoing surveys (Stark et al. in prep., Stern et al. in prep). The redshifts shown in the histogram have primarily been derived from the red-shifted Lyman-Alpha emission line (typically, asymmetric line profiles are seen in the spectra, which add confidence to their high redshift). The histogram shown in Figure 3 (left) only includes the highest quality targets from these surveys.

Table 1 shows the coordinates and redshifts for the overdensity centered on the histogram bin that contains HDF850.1. The redshifts are typically accurate to the third digit. 12 sources (13 including HDF850.1) are distributed over a narrow redshift range between 5.183<z<5.213 ($\Delta z = 0.03$). This corresponds to a difference in comoving radial distance of 15.7 Mpc (i.e. a proper distance of 2.5 Mpc). If one assumes spherical symmetry for the overdensity this distance would correspond to a projected extent of 6.6 arcmin, which is close to what is observed (Fig. 3, right).

We note that HDF 850.1 and the quasar are amongst the lowest redshift galaxies in the overdensity. One possible explanation is that Lyman Alpha emission lines (used to derive the redshifts of the remaining galaxies shown here) tend to be redshifted from the systemic velocity due to systematic absorption on the blue side of the line at these redshifts (the redshift determination of HDF850.1 through molecular emission lines is not affected by this bias).

**Table 1:** *List of sources in galaxy overdensity around HDF850.1.* The last column shows the redshift reference (Stern: Stern et al., in prep., Stark: Stark et al., in prep.)

| RA(2000.0) | DEC(2000.0) | redshift z | Reference |
|---|---|---|---|
| 12:36:00.0 | 62:12:26.1 | 5.199 | Stern |
| 12:36:26.5 | 62:12:07.4 | 5.200 | Stark |
| 12:36:37.5 | 62:12:36.0 | 5.185 | Stark |
| 12:36:39.8 | 62:09:49.1 | 5.187 | Stark |
| 12:36:48.0 | 62:09:41.4 | 5.186 | 18 |
| 12:36:49.2 | 62:15:38.6 | 5.189 | Stern, Stark |
| 12:36:52.0 | 62:12:25.8 | 5.183 | this work |
| 12:36:55.4 | 62:15:48.8 | 5.190 | Stark |
| 12:36:55.5 | 62:15:32.8 | 5.191 | Stern, Stark |
| 12:37:03.3 | 62:13:31.5 | 5.213 | Stern, Stark |
| 12:37:09.9 | 62:15:31.1 | 5.191 | Stark |
| 12:37:11.1 | 62:16:38.6 | 5.208 | Stern |
| 12:37:15.6 | 62:16:23.6 | 5.189 | Stern |

## 2. Spectral Energy Distribution of HDF 850.1

The observations presented here provide new constraints on the spectral energy distribution (SED) of HDF850.1. Figure S1 below summarizes all measurements (including data from the literature[2,3,6,39]) and includes new deep Herschel data of the region at 100, 160, 250, 350 and 500 microns[40]. For the Herschel bands we plot the measured flux density at the position of the source (0.18, 0.22, 5.0, 4.8 and 6.0 mJy, respectively) and add to this 2 times the noise (σ=0.5, 1.1, 3.2, 4.0, 4.0 mJy in the region around HDF850.1, respectively) as upper limits. We note that even though HDF850.1 is forming stars at a high rate, it is barely detected in the deep Herschel observations. The measurements done with the Spitzer Space telescope (shown in green) are blended with the bright foreground elliptical and thus only upper limits can be derived for HDF850.1[39]. The full line shows our best-fit SED using a simple model that provides a consistent energy balance between the attenuated stellar emission and the dust emission[41,42]. Our modeling shows that a minimum dust attenuation (defined as the difference between the observed and intrinsic magnitude at a given wavelength, $A_\lambda = m_\lambda[obs] - m_\lambda[model]$) of $A_{1500Å} \sim 5$ mag in the far-UV and $A_{5500Å} \sim 2$ mag in the V-band is required to be consistent with the Spitzer/IRAC upper limits, which sample the rest-frame optical. The only other high-redshift source with similar properties (no detection in rest-frame UV/optical) is GN10[43]. We note that an Arp220-like SED (shown as a dashed line) would result in even higher extinction values. An M82-like template (not shown), on the other hand, would be too bright in the rest-frame optical with respect to the observed upper limits. The emission by dust is described by using multiple components, including polycyclic aromatic hydrocarbon (PAH) features and a hot dust component in the mid-infrared, and dust in thermal equilibrium in the far-infrared, described by modified black bodies with an emissivity index of β=2.5. Lower values of β are difficult to reconcile with the observed steep drop in the Raleigh-Jeans part of the spectrum. This results in a total infrared luminosity of $L_{IR} = 8.7 \pm 1 \times 10^{12}$ $L_{sun}$ (FIR[42-122 μm] luminosity: $L_{FIR} = 6.5 \pm 1 \times 10^{12}$ $L_{sun}$) and an average dust temperature of T=35±5 K. The implied total dust mass is $M_{dust} = 2.75 \pm 0.5 \times 10^8$ $M_{sun}$, with a resulting molecular gas-to-dust ratio of 130±30(α/0.8), similar to what is found in local galaxies[44] within the uncertainties. The implied star formation rate is 850±100 $M_{sun}$ yr$^{-1}$ (assuming a Chabrier stellar initial mass function). We note that the absence of reliable measurements in the rest-frame infrared around the peak of the dust emission make it difficult to obtain an accurate redshift based on photometry alone, in particular if the FIR measurements have significant uncertainties[5].

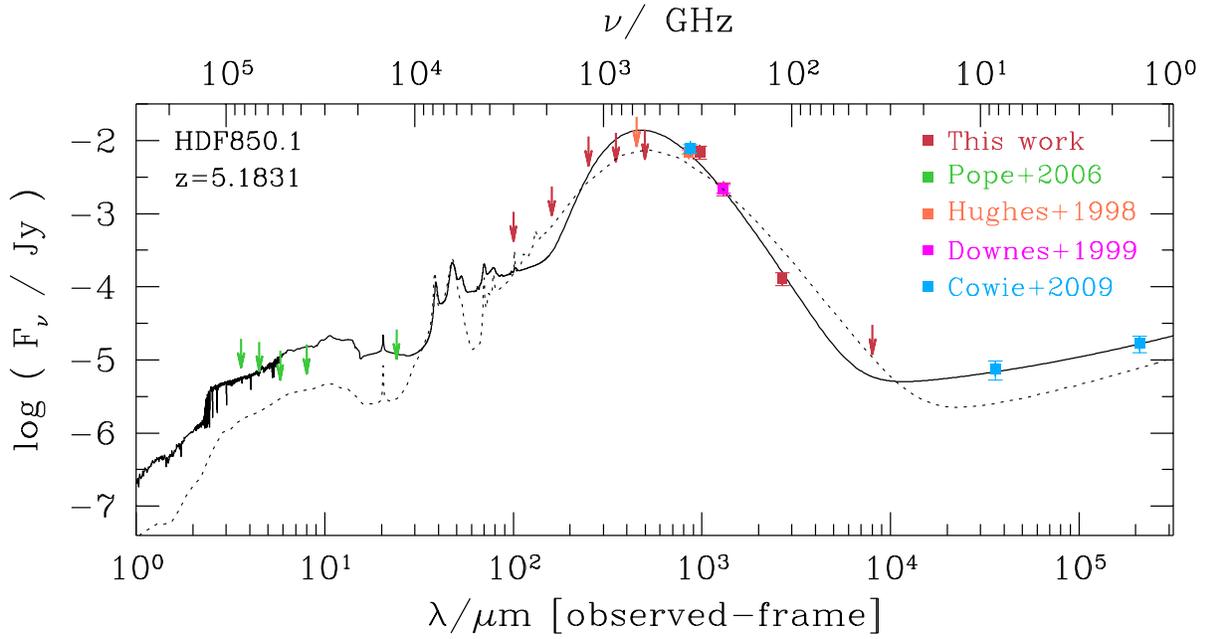

**Figure S1:** Spectral energy distribution (SED) of HDF850.1, including all available measurements from the literature and this work. The full line is our best-fit model (see text for details) – the dashed line is the scaled SED of Arp220 that is typically used to fit SEDs of high-redshift starburst galaxies. See Supplementary Information for details.

## 3. Stacking of fainter lines in the frequency scan

Our full frequency scan covers rest frequencies of 490-710 GHz for the redshift of HDF850.1. CO is by far the brightest emission line in this range, and we do not expect to detect the next brightest set of high-density tracers individually. We have however attempted to obtain a statistical detection of these lines through stacking. Figure S2 shows the result of the stacking exercise for the high-density tracers HCN(6-5), HCN(7-6), HCN(8-7), HNC(6-5), HNC(7-6), HCO+(6-5), HCO+(7-6), CS(11-10), CS(12-11), CS(13-12), CS(14-13), $HOC^+$(6-5), HOC+(7-6) that are covered in our frequency scan. No emission is detected in this stack with a 1σ flux limit of 0.06 Jy km s$^{-1}$ (assuming a linewidth of 400 km s$^{-1}$). This non-detection is consistent with the few detections of dense molecular gas tracers in high-redshift objects that are significantly fainter than the CO emission[45-54]. Also, no stacked detection was achieved using different combinations of high density tracers (e.g. stacking only on HCN, HNC and HCO+, or including additional species such as CN, CCH, and HC3N) or stacking on the CO isotopomers $^{13}$CO and C$^{18}$O.

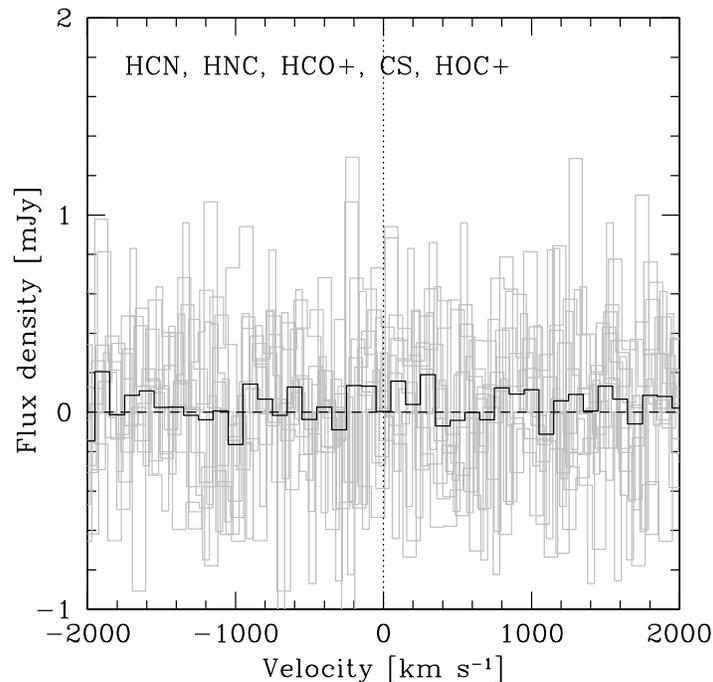

**Figure S2:** Spectral line stack for the high-density tracers (HCN, HNC, HCO$^+$, CS, HOC$^+$) covered by our frequency scan. The grey lines show the individual spectra before stacking whereas the black line shows the stacked result. No emission is detection in these lines (nor in any other combination of lines, see text).

**Additional References:**